\documentstyle[12pt]{article}
\begin{document}
\begin{center}
{\bf A COHERENT STATE ASSOCIATED WITH \\ SHAPE-INVARIANT
POTENTIALS}\footnote[1]{Talk presented by T.F. at the XIIth Workshop
on Geometric Methods in Physics, Bialowieza, Poland, July 1-7, 1993}
\footnote[2]{November, 1993. YITP/K-1041, RCNP-065}
\vskip1cm
{\footnotesize
T. FUKUI

{\it Yukawa Institute for Theoretical Physics,\\
Kyoto University, Kyoto 606-01, Japan}
\vskip0.5cm
and
\vskip0.5cm
N. AIZAWA

{\it Research Center for Nuclear Physics,\\
Osaka University, Ibaraki, Osaka 567, Japan}
}
\end{center}
\vskip0.5cm
\begin{abstract}
An algebraic treatment of shape-invariant potentials is discussed.
By introducing an operator which reparametrizes wave functions, the
shape-invariance condition
can be related to a generalized Heisenberg-Weyl algebra.
It is shown that this makes it possible to define a coherent state
associated with the shape-invariant potentials.
\end{abstract}
\vskip0.5cm\noindent{\bf 1. Introduction}\vskip0.5cm

Coherent states play an important role in physics today.$^1$
The original coherent state is closely related to Heisenberg-Weyl
group and this property has been extended to a number of Lie groups,
and the term coherent states is now applied to many objects.
Recently, a ``coherent state" is proposed$^2$ from a quite different
point of view. It is based on the idea of the reparametrization
invariance property of exactly solved potentials, called
shape-invariance$^{3, 4}$. This property is introduced by the use of
supersymmetry in quantum mechanics.$^{5}$
The key idea of constructing a coherent state
is to introduce an operator denoting the
reparametrization, which makes it possible to express
the shape-invariance condition as a commutation relation.
In this note, we construct a coherent state associated with
shape-invariant potentials based on this commutation relation.

\vskip0.5cm\noindent{\bf 2. Shape-Invariance Revisited}\vskip0.5cm
Let us first consider the following operators
\begin{equation}
D_n\equiv\frac{1}{\sqrt{2}}\left(W_n(x)+\frac{d}{dx}\right)
\end{equation}
with the property
\begin{equation}
D_nD_n^\dagger =D_{n+1}^\dagger D_{n+1}+R_{n+1}\quad (n=0,1,2,\cdots ),
\end{equation}
where $W_n(x)$ is so-called superpotential and
$R_{n+1}$ is a constant independent of $x$.
They are, in a sense, a generalization of the boson
creation and annihilation
operators. In terms of the superpotentials this relation is given by
\begin{equation}
W_n^2+W_n'=W_{n+1}^2-W_{n+1}'+2R_{n+1}\quad
(n=0,1,2,\cdots ),
\end{equation}
where dash denotes the derivative with respect to $x$.
This infinite chain of differential equations reduce to only one
equation if we adopt an anzats
\begin{equation}
W_n(x)=W(x,a_n), \quad
a_n\equiv\underbrace{f(f(\cdots f(}_{\mbox{$n$ times}}a_0)\cdots )).
\end{equation}
In fact, substituting this into Eq.3, we can easily confirm
the following statement: If the relation
\begin{equation}
W^2(x,a)+W'(x,a)=W^2(x,f(a))-W'(x,f(a))+2R(f(a)),
\end{equation}
holds identically with respect to $a$, the set of Eq.3
do not depend on $n$.
Now let us denote $D(a_n)\equiv D_n$. Then we have the relation
\begin{equation}
D(a)D^\dagger (a)=D^\dagger (f(a))D(f(a))+R(f(a))
\end{equation}
as the shape-invariance condition. Note that this resembles that
of the harmonic oscillator commutation relation.
However, because of the difference in parameter dependence
among operators, the above relation cannot be expressed naively
as a form a commutation relation.
Typical solutions of Eq.5 are as follows:
1) $f(a)=a-1, W(x,a)$ consists of finite power series of $a$.
In this case, we have six types of potentials classified by
Infeld and Hull.$^3$
2) $f(a)=qa, W(x,a)=aW(ax)$. This corresponds to the self-similar
potential discovered by Shabat and Spiridonov.$^4$
For such shape-invariant potentials we can calculate the
eigenvalues and corresponding eigenstates
in an algebraic way. To see this, let us consider
the following sequence of hamiltonians
\begin{eqnarray}
& &H_0\equiv D^\dagger (a_0)D(a_0)+R(a_0)\nonumber\\
& &H_1\equiv D(a_0)D^\dagger (a_0)+R(a_0)=D^\dagger (a_1)D(a_1)+
R(a_0)+R(a_1)\nonumber\\
& &\hskip5cm\vdots\nonumber\\
& &H_{n+1}\equiv D(a_n)D^\dagger (a_n)+\sum_{k=0}^{n}R(a_k)
=D^\dagger (a_{n+1})D(a_{n+1})+\sum_{k=0}^{n+1}R(a_k)\nonumber\\
& &\hskip5cm\vdots\nonumber
\end{eqnarray}
For these hamiltonians we can see the following two properties
provided that there exist normalized 0-eigenvalue states
satisfying $D(a_n)|\psi_0(a_n)\rangle =0\quad (n=0,1,2,...)$:
1) $D^\dagger (a_n)D(a_n)$ and $D(a_n)D^\dagger (a_n)$ are superpartners
each other, i.e., they have the same eigenvalues except for the
0-eigenvalue of the former. 2) The lowest eigenvalue of $H_{n+1}$
is $\sum_{k=0}^{n+1}R(a_k)$ since $D^\dagger (a_{n+1})D(a_{n+1})$
has a 0-eigenvalue.
Combining these two properties, we can conclude that $n$th eigenvalue
of $H_0$ and corresponding eigenstate are given by
\begin{eqnarray}
& &E_n(a_0)=\sum_{k=0}^nR(a_k),\nonumber\\
& &|\psi_n(a_0)\rangle \propto D^\dagger (a_0)D^\dagger (a_1)\cdots
D^\dagger (a_{n-1})|\psi_0(a_n)\rangle .
\end{eqnarray}
Therefore, once we know two important functions of $a$, i.e.,
$R(a)$ and $f(a)$, we can calculate eigenvalues and
eigenstates in an algebraic way.

\vskip0.5cm\noindent{\bf 3. A Coherent State Associated with
Shape-Invariant Potentials}\vskip0.5cm
It should be noted again that the shape-invariance condition Eq.6
can be considered as a generalization of the oscillator
commutation relation. To stress the analogy between them,
we introduce formally the operator
$T$ defined by
\begin{equation}
T|\phi (x,a)\rangle =|\phi (x,a_1)\rangle ,\quad a_1=f(a),
\end{equation}
which denotes the reparametrization of the parameter $a$.
Using this operator, we define the following operators
\begin{equation}
A_+(a)\equiv D^\dagger (a) T, \quad A_- (a)\equiv T^{-1}D(a) .
\end{equation}
Then we achieve the following expression of the shape-invariance
condition in the form of a commutation relation
\begin{equation}
[A_-(a), A_+(a)]=R(a).
\end{equation}
However, it is not closed in general because of the existence of
$T$ in $A$ operators.
{}From now on, we assume that $|\psi_0(a_n)\rangle \quad (n=0,1,...)$
are all normalizable eigenstates, i.e., $H_0$ has infinite number of
bound states. After some calculation we get the expression of
the normalized eigenstate of Eq.7
\begin{equation}
|\psi_n(a_0)\rangle =\frac{1}{\sqrt{[n]_0!}}\{A_+(a_0)\}^n|\psi_0(a_0)
\rangle ,
\end{equation}
where
\begin{eqnarray}
& &[n]_k\equiv R(a_{k+1})+R(a_{k+2})+\cdots +R(a_{k+n}) ,\quad
\widehat{[n]}_k\equiv [n]_kT ,\nonumber\\
& &[n]_k!\equiv\widehat{[n]}_k\widehat{[n-1]}_k\cdots\widehat{[1]}_k
\cdot T^{-n} .
\end{eqnarray}
The appearance of $T$ in $\widehat{[n]}_k$ reflects the
non-commutative character between $R(a)$ and $A_\pm(a)$.
This expression can be considered as a generalization of the expression
of the eigenstate of the usual harmonic oscillator, i.e., if we replace
$[n]_0$ and $A_+$ by the natural number and boson creation operator,
respectively, Eq.11 reduces to the well-known formula.

Now let us define a ``coherent state" associated with the commutation
relation Eq.10. Here coherent state means the eigenstate of the
``annihilation" operator $A_-(a)$.
For this purpose, we first define the generalized exponential
function
\begin{equation}
\exp_k(x)\equiv\sum_{n=0}^\infty\frac{1}{[n]_k!}x^n ,
\end{equation}
using Eq.12. Next, we define the state
\begin{eqnarray}
|z,a_0)&\equiv&\exp_0\{zA_+(a_0)\}|\psi_0(a_0)\rangle\nonumber\\
&=&\sum_{n=0}^\infty\frac{1}{\sqrt{[n]_0!}}z^n|\psi_n(a_0)\rangle ,
\end{eqnarray}
which is, of course, a generalized definition of the usual
boson coherent state.
{}From the direct calculation, we can easily confirm the relation
$A_-(a_0)|z,a_0)=z|z,a_0)$.

\vskip0.5cm\noindent{\bf 4. Examples with} $\it f(a)=a-1$\vskip0.5cm
Let us concretely calculate the newly defined coherent state Eq.14
for typical shape-invariant potentials.
We follow the classification of Infeld and Hull in Ref.3
and confine our attention to systems with only bound states.
Constants $a,b,c,d$ in this reference correspond to
$\alpha ,\beta ,\gamma ,\delta$, respectively.

\noindent (I) Types (C) and (D).$\quad$
These are the simplest cases among the shape-invariant potentials.
$W$ and $R$ in Eq.5 are given by
\begin{eqnarray}
&&W(x,a)=\left\{\begin{array}{ll}(a+\delta )/x+\beta x /2,&
\quad\mbox{for (C)}\\ \beta x+\delta ,&\quad
\mbox{for (D)}\end{array}\right.\nonumber\\&&R(a)=\beta ,
\end{eqnarray}
where $\beta$ and $\delta$ are some real constants.
In these cases we can set $R(a)=1$ without loss of generality. Then
$$
[n]_k=n,\quad [n]_k!=n!
$$
are independent of $k$, namely, they are the usual natural numbers and
factorial, respectively.
The coherent state becomes, therefore,
\begin{eqnarray}
&&|z, a_0)=\sum_{n=0}^\infty{z^n\over\sqrt{n!}}|{\psi_n(a_0)\rangle},
\nonumber\\&&{\cal N}_z(a_0)=\exp(|z|^2) ,
\end{eqnarray}
where ${\cal N}_z(a_0)=(z,a_0|z,a_0)$ is a normalization of the
coherent state, and from now on we denote the normalized coherent state
as $|z,a_0\rangle ={\cal N}^{-1/2}_z|z,a_0)$.
For the measure
\begin{equation}
d\mu (z)=\frac{1}{\pi}d Re z d Im z ,
\end{equation}
we have the following completeness relation
\begin{eqnarray}
& &\int d\mu (z)|{z,a_0}\rangle\langle z,a_0| \nonumber\\
& &=\sum_{m,n=0}^\infty\frac{1}{\sqrt{m!n!}}|\psi_m(a_0)\rangle
\langle\psi_n(a_0)|\frac{1}{\pi}\int d Re zd Im z
\exp (-|z|^2)\bar z^mz^n \nonumber\\
& &=\sum_{n=0}^\infty |\psi_n(a_k)\rangle\langle\psi_n(a_k)|. \nonumber
\end{eqnarray}

Next we briefly mention the coordinate representation of
the coherent state for these simplest cases.
Since the potential of type (D)
is that of the harmonic oscillator and it is almost trivial,
we restrict ourselves to type (C).
Since the coherent state is an eigenstate of $A_-$,
the relation
\begin{equation}
\{W(x,a_0)+d/dx\}\langle x|z,a_0)=\sqrt{2}zT\langle x|z, a_0)
\end{equation}
holds. The solution of this equation is given by
\begin{equation}
\langle x|z,a_0)=x^{-(a_0+\delta)}e^{-(\beta /4) x^2}
e^{zx^2T/\sqrt{2}}C(z,a_0),
\end{equation}
where $C(z,a_0)$ is independent of $x$.
On the other hand, by noting $D(a_0)|\psi (a_0)\rangle =0$,
the ground state in $x$-representation is given by
\begin{eqnarray}
&&\langle x|\psi_0(a_0)\rangle ={\cal N}^{-1/2}(a_0)
x^{-(a_0+\delta )}e^{-(\beta /4)x^2}, \nonumber\\
&&{\cal N}(a_0)=\int_{-\infty}^\infty dx x^{-2(a_0+\delta )}
e^{-(\beta /2)x^2}.
\end{eqnarray}
By using the property $\langle\psi_0(a_0)|z, a_0)=1$, we have
\begin{eqnarray}
1&&=\int_{-\infty}^\infty\ dx{\cal N}^{-1/2}(a_0)
x^{-(a_0+\delta )}e^{-(\beta /4)x^2}\cdot x^{-(a_0+\delta )}
e^{-(\beta /4)x^2}e^{zx^2T/\sqrt{2}}C(z,a_0) \nonumber\\
&&={\cal N}^{-1/2}(a_0)\sum_{n=0}^\infty\frac{1}{n!}\left(
\frac{z}{\sqrt{2}}\right)^n C(z,a_n){\cal N}(a_n)\nonumber\\
&&={\cal N}^{-1/2}(a_0)e^{(zT/\sqrt{2})}C(z, a_0)
{\cal N}(a_0).\nonumber
\end{eqnarray}
Solving this with respect to $C(z,a_0)$, we end up with
\begin{equation}
\langle x|z, a_0)=x^{-(a_0+\delta)}e^{-(\beta /4)x^2}e^{zx^2T/\sqrt{2}}
{\cal N}^{-1}(a_0)e^{-zT/\sqrt{2}}{\cal N}^{1/2}(a_0).
\end{equation}

\noindent (II) Types (A) and (B).$\quad$
$W$ and $R$ are given by
\begin{eqnarray}
&&W(x,a)=\left\{\begin{array}{ll}
\alpha (a+\gamma )\cot\alpha (x+p)+
\delta /\sin\alpha (x+p), &\quad\mbox{for (A)}\\
i\alpha (a+\gamma )+\delta\exp (-i\alpha x) ,&\quad\mbox{for (B)}
\end{array}\right.\nonumber\\
&&R(a)=-\alpha^2(a+\gamma +\frac{1}{2}),
\end{eqnarray}
where $\alpha$ is in general a real or a pure imaginary constant
for type (A), while it is pure imaginary for type (B),
and  $\gamma ,\delta$ and $p$ are real constants.
However, as we assume that the hamiltonian under consideration
has only bound states,
we confine our attention to the type (A) with real $\alpha$.
Note that in $W$ and $R$, $\gamma$ always appears as a form
$a+\gamma$. Therefore,
we can put $\gamma=0$ without loss of generality.
Furthermore, $R$ can reduce to $R(a)=-(a+1/2)$ by replacing
$A_\pm\rightarrow A_\pm/\alpha$. Finally, we can fix the various
parameters by choosing $a_0$ properly. Among them,
for example, if we choose $a_0=-1/2$, we have a simple expression
for the coherent state Eq.14 as follows; first we have
$$
[n]_k={1\over 2}n(2k+n+1) ,\quad
[n]_k!={1\over 2^n}n!{(2k+2n)!\over (2k+n)!}.
$$
It should be noted the factorial depends on $k$, namely, on $a$.
For $k=0$, we have $[n]_0!=(2n)!/{2^n}$ and
\begin{eqnarray}
&&|z,a_0) =\sum_{n=0}^\infty
\frac{1}{\sqrt{(2n)!}}z^n|\psi_n(a_0)\rangle ,\nonumber\\
&&{\cal N}_z(a_0)=\sum_{n=0}^\infty\frac{|z|^{2n}}{(2n)!}=\cosh (|z|) ,
\end{eqnarray}
where we replace $\sqrt{2}z\rightarrow z$. For the measure
\begin{equation}
d\mu_0(z)={1\over 2\pi}
{\cal N}_z(a_0){\exp (-|z|)\over |z|}d Re zd Im z,
\end{equation}
we have a similar completeness relation in example (I).

For other examples, see Ref.2.

\vskip0.5cm\noindent{\bf 5. An Example with} $f(a)=qa$\vskip0.5cm

Before calculating the coherent state Eq.14, we must confirm that
there really appears the q-oscillator algebra
since the shape-invariance
condition Eq.10 is here represented by the usual oscillator-like
commutation relation.
By substituting $f(a)=qa$ and $W(x,a)=aW(ax)$ into Eq.5, we have
\begin{equation}
W^2(x)+\frac{dW(x)}{dx}=q^2W^2(qx)-q\frac{dW(qx)}{dx}
+\frac{2R(f(a))}{a^2} .
\end{equation}
As previously mentioned, this equation should hold identically
with respect to $a$. The last term should be, therefore, a constant,
denoted here by $\gamma (q) (>0)$:
\begin{equation}
R(a)=\frac{\gamma (q)}{2q^2}a^2 .
\end{equation}
Hereafter, we set $\gamma (q)=2$ for simplicity.
The commutation relation denoting the shape-invariance Eq.10
is then given by
$[A_-(a), A_+(a)]=a^2/q^2$ .
Though this seems oscillator-like algebra, it is not closed since
$A$ operators and $a$ are not commutative.
It is possible, however, to get a closed relation
as follows: Let us introduce modified $A$-operators
\begin{equation}
A_{q+}(a)\equiv \frac{1}{a}A_+(a) ,\quad
A_{q-}(a)\equiv A_-(a)\frac{1}{a}.
\end{equation}
Then, the above commutataion relation is rewritten as
\begin{equation}
A_{q-}(a)A_{q+}(a)-q^2A_{q+}(a)A_{q-}(a)=1 ,
\end{equation}
which is essentially  equivalent to the one derived by
Spiridonov.$^4$

Next, let us calculate $[n]_0$ and $[n]_0!$ in Eq.12.
By definition, we have $a_k=q^ka_0$ and therefore $R(a_k)=q^{2(k-1)}
a_0^2$. Then we have
\begin{equation}
[n]_0=[n]_q\cdot a_0^2 ,\quad
[n]_0!=[n]_q!\cdot a_0^2a_1^2\cdots a_{n-1}^2 ,
\end{equation}
where $[n]_q\equiv (1-q^{2n})/(1-q^2)$ is a q-deformed $n$ and
$[n]_q!\equiv [n]_q[n-1]_q\cdots [1]_q$ is a q-deformed factorial.

Before we calculate the coherent state Eq.14, recall that we define
this state based on the commutation relation Eq.10.
What is important here is that this relation is invariant under
the transformation
\begin{equation}
A_{g+}(a)=g(a)A_+(a),\quad
A_{g-}(a)=A_-(a)\frac{1}{g(a)},
\end{equation}
where $g(a)$ is an arbitrary function of $a$. Using this property,
we can immediately define a coherent state which is the eigenstate
of $A_{g-}$ as follows;
\begin{equation}
A_{g-}|z,a_0)_g =z|z,a_0)_g
\end{equation}
with
\begin{equation}
|z,a_0)_g=\exp_0\{zA_{g+}(a_0)\}
|\psi_0(a_0)\rangle .
\end{equation}
Now let us choose $g(a)=a$. Then we have
\begin{equation}
A_{g-}(a)=A_{q-}(a), \quad
A_{g+}(a)=a^2A_{q+}(a) .
\end{equation}
By the use of Eq.29, the exponent in Eq.32 is calculated as follows;
\begin{eqnarray}
\exp_0\{zA_{g+}(a_0)\}&=&\sum_{n=0}^\infty
\frac{1}{[n]_q!\cdot a_0^2a_1^2\cdots a_{n-1}^2}
\{za_0^2A_{q+}(a_0)\}^n\nonumber\\
&=&\exp_q\{zA_{q+}(a_0)\} .
\end{eqnarray}
where $\exp_q(x)\equiv\sum_{n=0}^\infty x^n/[n]_q!$ is a q-deformed
exponential function.
Therefore, we conclude that a coherent state associated with
the shape-invariant potentials naturally leads to the q-coherent
state$^6$ in case of the self-similar potentials.

\vskip0.5cm\noindent{\bf 6. Summary}\vskip0.5cm

By introducing an operator $T$ denoting a reparametrization
of $a$, we have represented the shape-invariance condition as a form
of a commutation relation, and based on it we have defined a
coherent state associated with shape-invariant potentials.
It is shown that in cases of the usual harmonic oscillator and
recently found self-similar potentials, it reduces to the usual
and the q-deformed coherent state, respectively.
We expect this state should play a similar role as the usual and
generalized coherent states have been playing in various fields
in modern physics.

\vskip0.5cm\noindent{\bf 7. Acknowledgments}\vskip0.5cm

This work is supported in part by Grant-in-Aid from the Ministry
of Education, Science and Culture.
\vskip0.5cm\noindent{\bf 8. References}\vskip0.5cm
\begin{enumerate}\itemsep=0cm\parsep=0cm
\item
For reviews of coherent states, see, e.g.,
J. R. Klauder, in {\it Coherent States.
Applications in Physics and Mathematical Physics}, eds. J. R. Klauder
and B. S. Skagerstam (World Scientific, Singerpore, 1985):
A. Perelomov,{\it Generalized Coherent States and Their Applications}
(Springer-Verlag, Berlin, 1986).
\item
T. Fukui and N. Aizawa, Phys. Lett. {\bf A180} (1993) 308:
T. Fukui, preprint YITP/K-1034.
\item
L. Infeld and T. E. Hull, Rev. Mod. Phys. {\bf 23} (1951)
21: L. E. Gendenstein, JETP Lett.{\bf 38} (1983) 356;
L. E. Gendenstein and I. V. Krive, Sov. Phys. Usp. {\bf 28} (1985) 645:
F. Cooper, J. N. Ginocchio and A. Khare, Phys. Rev.
{\bf D36} (1987) 2458:
G. L\'evai, J. Phys. {\bf A22}, 689 (1989); J. Phys.
{\bf A25}, L521 (1992):
A. Das and Wen-Jui Huang, Phys. Rev. {\bf D41} (1990) 3241:
R. De, R. Dutt and U. Sukhatme, J. Phys. {\bf A25} (1992) L843;
Phys. Rev. {\bf A46} (1992) 6869:
C. A. Singh and T. H. Devi, Phys. Lett. {\bf A171}
(1992) 249:
\item
A. Shabat, Inver. Prob. {\bf 8} (1992) 303:
V. Spiridonov, Phys. Rev. Lett. {\bf 69} (1992) 398:
S. Skorik and V. Spiridonov, Lett. Math. Phys.
{\bf 28} (1993) 59:
\item
E. Witten, Nucl. Phys. {\bf B185} (1981) 513: E. Gozzi, Phys. Lett.
{\bf B129} (1983) 432.
\item
L. C. Biedenharn, J. Phys. {\bf A22} (1989) L873:
R. W. Gray and C. A. Nelson, J. Phys. {\bf A23} (1990) L945:
A. J. Bracken, D. S. McAnally, R. B. Zhang and M. D. Gould, J. Phys.
{\bf A24} (1991) 1379.
\end{enumerate}

\end{document}